\documentclass[11pt]{article}
\usepackage[dvips]{graphicx}
\begin{document}
\begin{center}
{\LARGE\bf  Cosmology with a Nonlinear Born-Infeld Type Scalar
Field  }
 \vskip 0.15 in
H.Q.Lu{\footnote{$alberthq_-lu@hotmail.~com$}}\\
{\it Department of Physics, \\Shanghai University, Shanghai 200436, China}\\
  \vskip 0.5 in
\centerline{Abstract} \vskip 0.2 in
\begin{minipage}{5.5in}
{ \hspace*{15pt}\small
 Recent many physicists suggest that the dark energy in the
 universe might result from the Born-Infeld(B-I) type scalar field of string
 theory. The universe of B-I type scalar field with potential can undergo a
 phase of accelerating expansion. The corresponding equation of state parameter
 lies in the range of $\displaystyle -1<\omega<-\frac{1}{3}$. The equation of
 state parameter of B-I type scalar field without potential lies in the range
 of $0\leq\omega\leq1$. We find that weak energy condition and strong energy
 condition are violated for phantom B-I type scalar field. The equation of state parameter
 lies in the range of $\omega<-1$.
  \\
  {\bf Keywords:}Dark energy;~Born-Infeld type scalar field;~Phantom cosmology.\\
 {\bf PACS:}98.80.Cq}
\end{minipage}
\end{center}
\newpage
\section{Introduction}
 \hspace*{15pt} Evidence that the universe is undergoing a phase of accelerating
  expansion at the present epoch continues to grow. This not only can be inferred
   by accelerating dynamics in high redshift surveys of Ia type supernovae[1], but
   also now it is independently implied from seven cosmic microwave background
   experiments(including the latest WMAP)[2], The favoured explanation for this
   behavior is that the universe is presently dominated by some form of dark energy
   density, contribution up to 70\% of the critical energy density, with the remaining
   30\% comprised of clumpy baryonic and non-baryonic dark matter. One of the central
   questions in cosmology today is the origin of the dark energy. Many candidates for
   dark energy have been proposed so far to fit the current observations. Among these
   models, the most important one is a self-interacting scalar field with a potential and
  thereby acts as a negative pressure source, referred to as "quintessence"[3]. This
  paradigm has caused attentions because a wide class of models exhibits tracking behavior
  at late time, where the dynamics of the field becomes independent of its initial conditions
  in early universe. In principle, this may resolve the fine-tuning inherent problem in
  dark energy models purely based on a cosmological constant. The major difference among
  these models is that different model predicts dissimilar equation of state of the dark
  energy, thus different cosmology is predicted. Especially, for these models, the equations
  of state parameter are confined within the range of $\displaystyle -1<\omega=\frac{p}{q}<-\frac{1}{3}$,
       which can drive the conclusion of accelerating expansion of the universe.
       However, some analysis of the observation data hold that the range of the
       equation of state parameter may not always be greater than $-1$ . In fact, they can lie in
       the range of $-1.48<\omega<-0.72$ [4]. It is obvious that the equation of state of conventional quintessence
        models based on a scalar field with positive kinetic energy can not evolve into the
        range of $\omega<-1$, and therefore, some authors[5] investigated phantom
        field models that possess negative kinetic energy and can realize $\omega<-1$ in their
        evolution. It is true that the field theory with negative kinetic energy poses a
        challenge to the widely accepted energy condition and leads to a rapid vacuum decay[6],
         but it is still very important to study these models, in some sense that is
         phenomenologically interesting.
  \par On the other hand, the role of tachyon field in string
  theory in cosmology has been widely studied[5]. It shows that
  the tachyon can be described by a Born-Infeld (B-I)type
  lagrangian resulting from string theory. It is  clear that the
  lagrangian $\displaystyle\frac{1}{\eta}[1-\sqrt{1-\eta g^{\mu\nu}\varphi_{,~\mu}\varphi_{,~\nu}}]-u(\varphi)$
  is equivalent formally to the tachyon type lagrangian $-\displaystyle\frac{1}{\eta}\sqrt{1-g^{\mu\nu}\Phi_{,~\mu}\Phi_{,~\nu}}$
  with a potential $\left[\displaystyle\frac{1}{\eta}-u(\varphi)\right]$,
  where re-scale the scalar field as $\Phi=(\eta)^{1/2}\varphi$.
  In this paper we consider cosmology of B-I type scalar
  field. The paper is organized as follows: In sec.2, we consider
  B-I type Lagrangian of scalar field without a potential. We
  obtain $0<\omega<1$. In sec.3, the B-I type Lagrangian of
  scalar field with a potential is considered. We find that
  potential $u(\varphi)$ is greater than $\displaystyle\frac{1}{\eta}$, and the
  kinetic energy of B-I scalar field is smaller than
  $\displaystyle\frac{1}{3\eta}$, thus we obtain $-1<\omega<-1/3$. In sec.4, the phantom with B-I
  Lagrangian is considered. We find that weak energy condition and
  strong energy condition are violated for phantom B-I type scalar field with potential.
  The equation of state parameter lies in the range $\omega<-1$.
  Sec.5 is Summary.
\section{The Model with B-I Lagrangian $\displaystyle\frac{1}{\eta}[1-\sqrt{1-\eta g^{\mu\nu}\varphi_{,~\mu}\varphi_{,~\nu}}]$}
\hspace*{15pt} In 1934[7], Born and Infeld put forward  a theory
of non-linear electromagnetic field. The lagrangian density is
$$\displaystyle L_{BI}=b^2\left[1-\sqrt{1-(\frac{1}{2b^2})F_{\mu\nu}F^{\mu\nu}}~\right]\eqno(2-1)$$
The lagrangian density for a B-I type scalar field is
$$\displaystyle L_S=\frac{1}{\eta}\left[1-\sqrt{1-\eta g^{\mu\nu}\varphi_{,~\mu}\varphi_{,~\nu}}~\right]\eqno(2-2)$$
Eq.(2-2) is equivalent to the tochyon lagrangian
$[-V(\varphi)\sqrt{1-g^{\mu\nu}\varphi_{,~\mu}\varphi_{,~\nu}}+\Lambda]$
if $\displaystyle V(\varphi)=\frac{1}{\eta}$ and cosmological
constant $\displaystyle \Lambda=\frac{1}{\eta}$ ($\displaystyle
\frac{1}{\eta}$ is two times as "critical" kinetic energy of
$\varphi$ field). The lagrangian (2-2) possesses some interesting
characteristics, it is exceptional in the sense that shock waves
do not develop under smooth or continuous initial conditions and
because nonsingular scalar field solution can be generated[8].
When $\eta \rightarrow 0$, by Taylor expansion, Eq.(2-2)
approximates to the lagrangian of linear scalar field.
$$\lim_{\eta\rightarrow 0}L_S=\frac{1}{2}g^{\mu\nu}\varphi_{,~\mu}\varphi_{,~\nu}\eqno(2-3)$$
A quantum model of gravitation interacting with a lagrangian (2-2)
of B-I type scalar has been considered by us. We obtained the
Wheeler-Dewitt equation of B-I scalar field and found the wave
function of the universe. An inflationary universe, with the
largest possible vacuum energy and the largest interaction between
the particles of B-I scalar field[9], is predicted. Next we
consider classical cosmology. For the spatially homogeneous scalar
field, Eq.(2-2) becomes
$$L_S=\frac{1}{\eta}\left[1-\sqrt{1-\eta \dot{\varphi}^2}~\right]\eqno(2-4)$$
In the spatially flat Robertson-Walker metric
$ds^2=dt^2-a^2(t)(dx^2+d^2y+d^2z)$, Einstein equation
$G_{\mu\nu}=KT_{\mu\nu}$ can be written as
$$\left(\frac{\dot{a}}{a}\right)^2=\frac{K}{3}T^0_0\eqno(2-5)$$
$$2\frac{\ddot{a}}{a}+\left(\frac{\dot{a}}{a}\right)^2=KT^1_1=KT^2_2=KT^3_3\eqno(2-6)$$
Substituting Eq.(2-5) into Eq.(2-6), we get
$$\frac{\ddot{a}}{a}=-\frac{K}{3}(T^0_0-3T^1_1)\eqno(2-7)$$
where
$$T^\mu_\nu=\frac{g^{\mu\rho}\varphi_{,~\nu}\varphi_{,~\rho}}{\sqrt{1-\eta g^{\mu\nu}\varphi_{,~\mu}\varphi_{,~\nu}}}-\delta^\mu_\nu L_S \eqno(2-8)$$
The energy density $\rho_s=T^0_0$ and pressure $p_s=-T^i_i$ are
defined as following:
$$\rho_s=T^0_0=\frac{\dot{\varphi}^2}{1-\eta\dot{\varphi}^2}-L_s\eqno(2-9)$$
$$p_s=-T^i_i=\frac{1}{\eta}[1-\sqrt{1-\eta\dot{\varphi}^2}]\eqno(2-10)$$
where the upper index "." denotes the derivative with respect to
$t$.\\The equation of motion of scalar field $\varphi$ is
$$\frac{1}{\sqrt{-g}}\frac{\partial}{\partial x^\nu}\left[\frac{\sqrt{-g}g^{\mu\nu}\varphi_{,~\mu}}{\sqrt{1-\eta g^{\mu\nu}\varphi_{,~\mu}\varphi_{,~\nu}}}\right]=0\eqno(2-11)$$
The scalar field $\varphi$ only depends on $t$. From Eq.(2-11), we
can obtain
$$\dot{\varphi}=\frac{c}{\sqrt{a^6+\eta c^2}}\eqno(2-12)$$
where $c$ is integral constant. When $a(t)=0$, the kinetic energy
$\dot{\varphi}^2=\displaystyle\frac{1}{\eta}$ is critical maximum.
From Eqs.(2-5),(2-7),(2-9),(2-10),(2-12) we get
$$\left(\frac{\dot{a}}{a}\right)^2=\frac{K}{3\eta}[\sqrt{1+\eta c^2a^{-6}}-1]\eqno(2-13)$$
$$\ddot{a}=-\frac{K}{3\eta}\left[2-\frac{2a^6-\eta c^2}{a^3\sqrt{a^6+\eta c^2}}\right]\eqno(2-14)$$
From Eq.(2-14) we can find that $\ddot{a}$ is always smaller than
zero , no matter what the value of $a(t)$ is. When
$a\rightarrow\infty$, then $\ddot{a}\rightarrow 0$. It shows that
the universe starts with decelerated regime and gradually enters
the zero acceleration. From Eq.(2-13) we get
$$\dot{a}=\displaystyle\sqrt{\frac{K}{3\eta}\left[a^2\sqrt{1+\eta c^2 a^{-6}}-a^2~\right]}\eqno(2-15)$$
By Eq.(2-15), then we can find that the minimum of $a(t)$ is zero,
assume $a(t)\mid_{t=0}=0$. When $a(t)\ll 1$, Eq.(2-15) can be
approximated as
$$\sqrt{a}~\dot{a}\approx\sqrt{\frac{Kc}{3}\sqrt{\eta}}\eqno(2-16)$$
$$a^{3/2}\approx\frac{3}{2}\sqrt{\frac{Kc}{3}\sqrt{\eta}}~t\eqno(2-17)$$
$$a(t)\sim t^{2/3}\eqno(2-18)$$
For the energy density $\rho_\varphi$ (2-9) and pressure
$p_\varphi$ (2-10) of B-I scalar field, there is no violation of
the strong energy condition. In universe with B-I scalar field
without
potential, there is no a phase of accelerating expansion.\\
From Eqs.(2-9)(2-10)and (2-12), we have
$$\omega=\frac{p_\varphi}{\rho_\varphi}=\frac{a^3}{\sqrt{a^6+\eta c^2}}\eqno(2-19)$$
and can see
$$0\leq\omega\leq1\eqno(2-20)$$
In Sec.3, it is different that we find the universe accelerating
expansion.
\section{The Model
with Lagrangian of $\displaystyle\frac{1}{\eta}[1-\sqrt{1-\eta
g^{\mu\nu}\varphi_{,~\mu}\varphi_{,~\nu}}]-u(\varphi)$}
\hspace*{15pt} From Eq.(2-9) and lagrangian, we can get
$$T^0_0=\rho_\varphi=u-\frac{1}{\eta}+\frac{1}{\eta\sqrt{1-\eta \dot{\varphi}^2}}\eqno(3-1)$$
$$-T^i_i=p_\varphi=\frac{1}{\eta}[1-\sqrt{1-\eta\dot{\varphi}^2}]-u\eqno(3-2)$$
When $u>0$, so we can always find $\rho_\varphi>0$.\\
From Eqs.(3-1)and (3-2), we have
$$\rho_\varphi+3p_\varphi=\frac{2}{\eta}-2u+\frac{3\eta\dot{\varphi}^2-2}{\eta\sqrt{1-\eta\dot{\varphi}^2}}\eqno(3-3)$$
When potential is greater than $\displaystyle\frac{1}{\eta}$
($\displaystyle\frac{1}{\eta}$ is two times as "critical" kinetic
energy of $\varphi$ field). And the kinetic energy of $\varphi$
field evolves to region of
$\displaystyle\dot{\varphi}^2<\frac{2}{3\eta}$, we have
$\rho_\varphi+3p_\varphi<0$ from Eq.(3-3).\\
 The universe undergoes a phase of accelerating expansion.\\
 From Eqs.(3-1)and (3-2), we also can get
 $$p_\varphi+\rho_\varphi=\frac{\dot{\varphi}^2}{\sqrt{1-\eta \dot{\varphi}^2}}>0\eqno(3-4)$$
 Eq.(3-4) could be written as
 $$\omega=\frac{p_\varphi}{\rho_\varphi}>-1\eqno(3-5)$$
 when $\dot{\varphi}$ approximation zero
 $p_\varphi=-\rho_\varphi$. The universe is dominated by the
 potential. It will undergo inflation phase. In next section, we
 consider the case that the kinetic energy term is negative.

\section{The Model with Lagrangian of $\displaystyle\frac{1}{\eta}[1-\sqrt{1+\eta
g^{\mu\nu}\varphi_{,~\mu}\varphi_{,~\nu}}]-u(\varphi)$}
\hspace*{15pt} In this section we consider the case that the
kinetic energy term is negative.
$$L=\frac{1}{\eta}[1-\sqrt{1+\eta
g^{\mu\nu}\varphi_{,~\mu}\varphi_{,~\nu}}]-u(\varphi)\eqno(4-1)$$
The Energy-moment tensor is
$$\displaystyle T^\mu_\nu=-\frac{g^\mu_\nu \varphi_{,~\nu}\varphi_{,~\rho}}{\sqrt{1+\eta g^{\mu\nu}\varphi_{,~\mu}\varphi_{,~\nu}}}-\delta^\mu_\nu L\eqno(4-2)$$
From Eq.(4-2), we have
$$\rho=T^0_0=\frac{1}{\eta\sqrt{1+\eta \dot{\varphi}^2}}-\frac{1}{\eta}+u\eqno(4-3)$$
$$p=-T^i_i=\frac{1}{\eta}-\frac{\sqrt{1+\eta \dot{\varphi}^2}}{\eta}-u\eqno(4-4)$$
From Eqs.(4-3) and (4-4), we get
$$\rho+p=-\frac{\dot{\varphi}^2}{\sqrt{1+\eta \dot{\varphi}^2}}\eqno(4-5)$$
It is clear that the equation of static $\omega<-1$ is completely
confirmed by Eq.(4-5)and it accords with the recent analysis of
observation data. We also can get
$$\rho+3p=\frac{2}{\eta}\left[\frac{-2-3\eta \dot{\varphi}^2}{\sqrt{1+\eta \dot{\varphi}^2}}\right]-2u\eqno(4-6)$$
It is obvious that $\rho+3p<0$. Eq.(4-6)shows that the universe is
undergoing a phase of accelerating expansion. The model of phantom
B-I scalar field without potential $u(\varphi)$ is hard to
understand. In this model we can always find
$\displaystyle\rho=\frac{1}{\eta\sqrt{1+\eta\dot{\varphi}^2}}-\frac{1}{\eta}<0$
and
$\displaystyle\left(\frac{\dot{a}}{a}\right)^2=\frac{K}{3}\rho<0$.
It is unreasonable apparently. However, in the model of phantom
B-I scalar field with potential $u(\varphi)$, when $\displaystyle
u(\varphi)>\frac{1}{\eta}-\frac{1}{\eta\sqrt{1+\eta\dot{\varphi}^2}}$,
$\rho$ is always greater than zero. In phantom B-I scalar model
with a potential $u(\varphi)$, we also find the strong and weak
energy condition always failed from Eqs.(4-5)and (4-6).
\par We investigate the case of a specific simple example
$u=u_0=const$ and
$u_0-\displaystyle\frac{1}{\eta}=\frac{A}{\eta}(A>0)$. Eq.(4-3)
becomes
$$\displaystyle\rho=\frac{1}{\eta\sqrt{1+\eta\dot{\varphi}^2}}+
\frac{A}{\eta}\eqno(4-7)$$ It is clear that there is $\rho>0$ from
Eq.(4-7). Substituting Eq.(4-7) into Einstein equation, we have
$$\left(\frac{\dot{a}}{a}\right)^2=\frac{K}{3}\left[\frac{1}{\eta\sqrt{1+\eta\dot{\varphi}^2}}+\frac{A}{\eta}\right]\eqno(4-8)$$
The equation of motion of phantom field is
$$\frac{1}{\sqrt{-g}}\frac{\partial}{\partial x^\nu}\left[\frac{\sqrt{-g}g^{\mu\nu}\varphi_{,~\mu}}{\sqrt{1+\eta g^{\mu\nu}\varphi_{,~\mu}\varphi_{,~\nu}}}\right]=0\eqno(4-9)$$
The field $\varphi$ only depends on  $t$. We can obtain from
Eq.(4-9)
$$\dot{\varphi}=\frac{C}{\sqrt{a^6-\eta C^2}}\eqno(4-10)$$
where $C$ is integrate constant. Substituting Eq.(4-10) into
Eq.(4-8), the Eq.(4-8)becomes
$$\dot{a}=\displaystyle\sqrt{\frac{Ka^2}{3\eta}\left[\sqrt{1-\eta C^2a^{-6}}+A~\right]}\eqno(4-11)$$
The smallest $a_{\min}=(\eta C^2)^{1/6}$ from Eq.(4-11), the
universe is non-singular. When the universe scalar approximation
$a_{\min}$, Eq.(4-11) becomes
$$\dot{a}=\sqrt{\frac{KA}{3\eta}}~a\eqno(4-12)$$
$$a\sim exp\left(\displaystyle \sqrt{\frac{KA}{3\eta}}~t\right)\eqno(4-13)$$
When $a\rightarrow\infty$, Eq.(4-11) becomes
$$\dot{a}=\sqrt{\frac{K(A+1)}{3\eta}}~a\eqno(4-14)$$
$$a\sim exp\left(\displaystyle \sqrt{\frac{K(A+1)}{3\eta}}~t\right)\eqno(4-15)$$
In our phantom model with potential, the universe is always
undergoing a phase of inflation and gradually enters the more
accelerated expansion in late time.
\section{Summary}
\hspace*{15 pt} We considers cosmological solution of B-I type
scalar field without the potential and come to the conclusion that
the equation of state parameter $0<\omega<1$. However, in the dark
energy models of canonical B-I scalar field with potential, the
universe is undergoing a phase of accelerating expansion if the
potential rolls down to the minimum which is greater than
$\displaystyle\frac{1}{\eta}$ while scalar field evolves into
region of $\displaystyle\dot{\varphi}^2<\frac{2}{3\eta}$.
Correspondingly the equation of state parameter $\omega$ is always
greater than $-1$. This model admits  a late time attractor
solution that leads to an equation of state $\omega=-1$. The
lagrangian of B-I type scalar field with negative kinetic energy
also is considered by us. In the phantom B-I scalar model, the
universe is undergoing a phase of accelerating expansion and the
equation of state parameter $\omega$ is always smaller than $-1$.
It accords with the recent analysis of the observation that the
equation of state parameter of dark energy might be smaller
than $-1$. \\
{\bf Acknowledgment}\\
This work was partially supported by National Nature Science
Foundation of China.\\ \vskip 0.5in
 {\noindent\Large \bf References}
\small{
\begin{description}
\item {1.} {S.perlmutter et.al., Ap.J, 565(1999);\\J.L.Tonry et.al., astro-ph/0305008.}
\item {2.} {C.L.bennett et.al., astro-ph/0302207;\\
             A.Melhiori and L.Mersini, C.J.Odmann, and M.Thodden, astro-ph/0211522;\\
             D.N.Spergel et.al., astro-ph/0302209;\\
             N.W.Halverson et.al., Ap.J.568, 38(2002);\\
             C.B.Netterfield et.al., astro-ph/0104460;\\
             P.de Bernardis et.al., astro-ph/0105296;\\
             A.T.Lee et.al., astro-ph/0104460;\\
             R.Stompor, astro-ph/0105062. }
\item {3.} {J.A.Frieman and I.Waga, Phys.Rev.D 57, 4642(1998);\\
             R.R.Caldwell, R.Dave and P.J.Steinhardt, Phys.rev.Lett 80, 1582(1998).}
\item {4.} {R.R.Caldwell, Phys.Rev.Lett. B 545, 23(2002);\\
             V.Faraoni, Int.J.Mod.Phys. D 11, 471(2002);\\
             S.Nojiri and S.D.Odintsov, hep-th/0304131;hep-th/0306212;\\
             E.schulz and M.White, Phys.Rev.D 64, 043514(2001);\\
             T.Stachowiak and Szydllowski, hep-th/0307128;\\
             G.W.Gibbons, hep-th/0302199;\\
             A.Feinstein and S.Jhingan, hep-th/0304069.  }
\item {5.} {S.M.Carroll, M.Hoffman, M.Teodden, astro-th/0301273;\\
             Y.S.piao, R.G.Cai, X.M.Zhang and Y.Z.Zhang, hep-ph/0207143;\\
             J.G.Hao and X.Z.Li, hep-th/0305207;\\
             S.Mukohyama, Phys.Rev.D 66,024009(2002);\\
             T.Padmanabhan, Phys.Rev.D 66, 021301(2002);\\
             M.Sami and T.Padamanabhan, Phys.Rev.D67, 083509(2003);\\
             G.Shiu and I.Wasserman, Phys.Lett.B 541,6(2002);\\
             L.kofman and A.Linde, hep-th/020512;\\
             H.B.Benaoum, hep-th/0205140;\\
             L.Ishida and S.Uehara, hep-th/0206102;\\
             T.Chiba, astro-ph/0206298;\\
             T.Mehen and B.Wecht, hep-th/0206212;\\
             A.Sen, hep-th/0207105;\\
             N.Moeller and B.Zwiebach, JHEP 0210, 034(2002);\\
             J.M.Cline, H.Firouzjahi and P.Martineau. hep-th/0207156;\\
             S.Mukohyama, hep-th/0208094;\\
             P.Mukhopadhyay and A.Sen, hep-th/020814;\\
             T.Okunda and S.Sugimoto, hep-th/0208196;\\
             G.gibbons, K.Hashimoto and P.Yi, hep-th/0209034;\\
             M.R.Garousi, hep-th/0209068;\\
             B.Chen, M.Li and F.Lin, hep-th/0209222;\\
             J.Luson, hep-th/0209255;\\
             C.kin, H.B.Kim and O.K.Kwon, hep-th/0301142;\\
             J.M.Cline, H.Firouzjahi and P.Muetineau, hep-th/0207156;\\
             G.Felder, L.Kofman and A.Starobinsky, JHEP 0209, 026(2002);\\
             S.Mukohyama, hep-th/0208094;\\
             G.A.Diamandis, B.C.Georgalas, N.E.Mavromatos, E.Pantonopoulos, hep-th/0203241;\\
             G.A.Diamandis, B.C.Georgalas, N.E.Mavromatos, E.Pantonopoulos, I.Pappa, hep-th/0107124;\\
             M.C.Bento, O.Bertolami and A.A.Sen, hep-th/020812;\\
             H.Lee, et.al., hep-th/0210221;\\
             M.Sami, P.Chingangbam and T.Qureshi, hep-th/0301140;\\
             F.Leblond, A.W.Peet, hep-th/0305059;\\
             J.G.Hao and X.Z.Li, Phys.Rev.D 66, 087301(2002);\\
             X.Z.Li and X.H.Zhai, Phys.Rev.D 67, 067501(2003).}
\item {6.} {G.Felder, L.Kofman and A.Starobinsky, hep-th/0208019;\\
             G.W.Gibbons, hep-th/031117; hep-th/0302199;\\
             A.Frolov, L.Kofman and A.Starobinsky, hep-th/0204187;\\
             A.Sen, hep-th/0204143; hep-th/0209122; hep-th/0203211.}
\item {7.} {M.Born and Z.Infeld, Proc.Roy.Soc A 144, 425(1934).}
\item {8.} {H.P.de Oliveira, J.Math.Phys. 36, 2988(1995).}
\item {9.} {H.Q.Lu,T.Harko and K.S.cheng, Int.J.Modern.Phys.D 8,625(1999);\\
             H.Q.Lu et.al., Int.J.Theory.Phys.42, 837(2003). }
\end{description}}
\end{document}